# Rapid and Quantitative Chemical Exchange Saturation Transfer (CEST) Imaging with Magnetic Resonance Fingerprinting (MRF)

Ouri Cohen[1†], Shuning Huang[1†§], Michael T. McMahon[2,3], Matthew S. Rosen[1,4], Christian T. Farrar[1*]

[1]Athinoula A. Martinos Center for Biomedical Imaging, Department of Radiology, Massachusetts General Hospital and Harvard Medical School, Charlestown, MA, USA.
[2]F.M. Kirby Center for Functional Brain Imaging, Kennedy Krieger Institute, Baltimore, Maryland, USA.
[3]Department of Radiology, Johns Hopkins University School of Medicine, Baltimore, Maryland, USA.
[4]Department of Physics, Harvard University, Cambridge, MA, USA.

[†]These authors contributed equally to this work

**Grant sponsor:** National Institutes of Health; Grant number: R01CA203873.

**\*Correspondence to:** Christian T. Farrar, Athinoula A. Martinos Center for Biomedical Imaging, Department of Radiology, Massachusetts General Hospital, 149 13th Street, Suite 2301, Charlestown, MA, 02129, USA. Email: cfarrar@nmr.mgh.harvard.edu

**Present Address:** [§]Department of Biomedical Engineering, Texas A&M University, College Station, TX, USA.

**Submitted to Magnetic Resonance in Medicine**

**ABSTRACT**

**Purpose:** To develop a fast magnetic resonance fingerprinting (MRF) method for quantitative chemical exchange saturation transfer (CEST) imaging.

**Methods:** We implemented a CEST-MRF method to quantify the chemical exchange rate and volume fraction of the $N_\alpha$-amine protons of L-arginine (L-Arg) phantoms and the amide and semi-solid exchangeable protons of *in vivo* rat brain tissue. L-Arg phantoms were made with different concentrations (25-100 mM) and pH (pH 4-6). The MRF acquisition schedule varied the saturation power randomly for 30 iterations (phantom: 0-6 µT; *in vivo*: 0-4 µT) with a total acquisition time of ≤2 minutes. The signal trajectories were pattern-matched to a large dictionary of signal trajectories simulated using the Bloch-McConnell equations for different combinations of exchange rate, exchangeable proton volume fraction, and water T1 and T2* relaxation times.

**Results:** The chemical exchange rates of the $N_\alpha$-amine protons of L-Arg were significantly ($p<0.0001$) correlated with the rates measured with the Quantitation of Exchange using Saturation Power method. Similarly, the L-Arg concentrations determined using MRF were significantly ($p<0.0001$) correlated with the known concentrations. The pH dependence of the exchange rate was well fit ($R^2=0.9186$) by a base catalyzed exchange model. The amide proton exchange rate measured in rat brain cortex (36.3±12.9 Hz) was in good agreement with that measured previously with the Water Exchange spectroscopy method (28.6±7.4 Hz). The semi-solid proton volume fraction was elevated in white (11.2±1.7%) compared to gray (7.6±1.8%) matter brain regions in agreement with previous magnetization transfer studies.

**Conclusion:** CEST-MRF provides a method for fast, quantitative CEST imaging.

# INTRODUCTION

Chemical Exchange Saturation Transfer (CEST) MRI (1-3) uses selective radio-frequency (RF) pulses to saturate the magnetization of exchangeable protons on a variety of molecules and macromolecules, including proteins, which due to fast chemical exchange with bulk water results in a decreased water MRI signal. CEST has proven to be a powerful tool for imaging a wide range of disease states and pathologies. For example, the amide proton CEST contrast from endogenous proteins has been used to distinguish tumor progression from radiation necrosis in a gliosarcoma rodent model (4) and in clinical glioma patients (5,6), detect early response to temozolomide (7) and radiation therapy (8) in glioblastoma rodent models, evaluate tumor grade and cellularity of clinical glioma patients (9), distinguish benign and atypical meningiomas in clinical subjects (10), and detect changes in pH during stroke that may provide insight into the viability of the ischemic penumbra (11-13). In addition, a number of exogenous diamagnetic CEST imaging probes have been identified including lysine rich proteins (14,15), glucose (16-18), creatine (19,20), glycosaminoglycans (21), barbituric acid (22), thymidine analogs (23), iodinated compounds (24,25), imidazoles (26), salicylic acid analogs (27-29), and anthranillic (30) analogs. Glucose and iodinated CEST imaging probes are currently under clinical evaluation for monitoring tumor perfusion (31,32) and tumor acidosis (33), respectively. However, efficient methods for quantification of the chemical exchange rates and exchangeable proton volume fractions are needed to produce high quality pH and volume fraction maps required to move many of these studies forward.

In a traditional CEST experiment, the frequency offset of the RF saturation pulse is stepped across the water resonance to generate a CEST Z-spectrum of signal intensity as a function of saturation frequency offset. The Magnetization Transfer Ratio asymmetry ($MTR_{asym}$), or CEST contrast, is then calculated from the difference between the signal intensities (S) acquired with negative ($\omega^-$) and positive ($\omega^+$) frequency offsets from water as given by equation 1.

$$MTR_{asym} = \frac{S(\omega^-) - S(\omega^+)}{S_0} \qquad [\text{Eq. 1}]$$

However, CEST MRI suffers from several limitations including long image acquisition times and the qualitative nature of the CEST contrast, which depends on many factors, including the chemical exchange rate ($k_{ex}$), volume fraction of the exchangeable solute protons ($f_s$), water longitudinal relaxation rate ($R_{1w}$), RF saturation time ($t_{sat}$), and the RF saturation efficiency (α), which in turn depends on the saturation power ($B_1$), water transverse relaxation rate ($R_{2w}$), and $k_{ex}$, as given in equation 2 (34).

$$MTR_{asym} = \frac{\alpha \cdot f_s \cdot k_{ex}}{R_{1w} + \alpha \cdot f_s \cdot k_{ex}}\left(1 - e^{-(R_{1w} + \alpha \cdot f_s \cdot k_{ex})t_{sat}}\right) \qquad [\text{Eq. 2}]$$

$$\alpha = \frac{(\gamma B_1)^2}{(\gamma B_1)^2 + k_{ex}(k_{ex} + R_{2w})} \qquad f_s = \frac{[exchangeable\ proton]}{2 \cdot [H_2O]}$$

Analysis of the $MTR_{asym}$ is further complicated by the presence of Nuclear Overhauser Enhancement (NOE) effects attributed to aliphatic exchangeable protons between -2 to -5 ppm from water and very broad magnetization transfer (MT) effects due to semi-solid, macromolecular exchangeable protons centered at approximately -2.5 ppm. Clinical translation of CEST methods would therefore benefit greatly from the development of more quantitative, specific and rapid CEST methods.

A recently developed Magnetic Resonance Fingerprinting (MRF) method has been used for the rapid quantification of tissue T1 and T2 relaxation times (35) and for the multiparametric estimation of brain hemodynamic parameters such as cerebral blood flow (36). The MRF method varies the image acquisition parameters to generate unique signal trajectories for different quantitative tissue parameters. The experimental trajectories are then matched to a large dictionary of signal trajectories simulated using the Bloch equations for different combinations of tissue parameters. The MRF method allows for the simultaneous quantification of multiple parameters in a short acquisition time period. Here we extend the MRF approach by incorporating chemical exchange into the Bloch equation simulations, and report the use of a fast CEST-MRF method for generating quantitative exchange rate and proton volume fraction maps of $N_\alpha$-amine exchangeable protons of L-arginine phantoms with different concentrations (25-100 mM) and pH (pH 4-6), and of endogenous amide and semi-solid exchangeable protons of *in vivo* rat brain tissue.

## METHODS

### L-Arginine Phantoms

A set of phantoms was prepared with various L-arginine (L-Arg) concentrations by dissolving L-Arg (Sigma-Aldrich, St. Louis, MO) in pH 4 or pH 5 buffer (BDH, London, UK) at concentrations of 25, 50 or 100 mM. In addition, a set of phantoms with varying pH was prepared by titrating a 50 mM, pH 4 L-Arg solution with NaOH to a pH of 4.0, 4.5, 5.0, 5.5 or 6.0. The $N_\alpha$-amine of L-Arg has a chemical shift of +3 ppm with respect to the water resonance and has 3 equivalent exchangeable amine protons. The different L-Arg solutions were placed in 2 ml glass vials with sets of 3 vials placed into 50 ml Falcon tubes with 2% agarose gel surrounding the vials.

### Animal Preparation

All animal experiments and procedures were performed in accordance with the NIH Guide for the Care and Use of Laboratory Animals and were approved by the Institutional Animal Care and Use Committee of the Massachusetts General Hospital. Male Wistar rats (Charles River Labs, Wilmington, MA) were anesthetized with 1-2% isolflurane in 50:50 $O_2$:medical air mixture and placed prone on a home-built rat MRI cradle with ear and bite bars to secure the rat head. Respiration rate and temperature were monitored with a small animal physiological monitoring system (SA Instruments, Inc., Stony Brook, NY) and a body temperature of 37°C was maintained by blowing warm air in the bore of the magnet.

### Magnetic Resonance Imaging

Single-slice, single-shot CEST gradient Echo Planar Images (EPI) were acquired on a 4.7 T MRI scanner (Bruker Biospin, Billerica, MA). Phantom images were acquired with a 35 mm inner diameter birdcage volume coil (Bruker Biospin, Billerica, MA), while the *in vivo* rat images were acquired with a rat brain, 4-channel, phased array-receive coil (Bruker Biospin, Billerica, MA) and a 72 mm quadrature birdcage volume transmit coil (Bruker Biospin, Billerica, MA). The CEST-MRF acquisition schedule (shown schematically in Figure 1) was designed to keep the saturation pulse frequency offset fixed at the amine (L-Arg α-$NH_3$: +3 ppm) or amide (rat brain: +3.5 ppm) exchangeable proton frequency, and randomly vary the saturation power for 30 iterations, with amplitude between 0-6 µT for the phantom studies and 0-4 µT for the *in vivo* studies. For the phantom studies the image acqusition parameters were: saturation pulse length = 3000 ms, TE/TR=21/4000 ms, flip angle (FA) = 60°, matrix 100x100, field of view (FOV) = 30x30 mm, number of averages (NA) = 1. For the *in vivo* study the image acquisition parameters were: saturation pulse length = 2500 ms, TE/TR = 21/3500 ms, FA = 90°, matrix 80x80, FOV = 40x40 mm, NA = 1. The total MRF image acquisition time was 2 minutes for the phantom study and 1.75 minutes for the *in vivo* study.

For the L-Arg phantoms, the amine proton chemical exchange rates were independently measured with the QUantification of Exchange using Saturation Power (QUESP) MRI method (37). Single-shot QUESP-EPI images were acquired at saturation frequency offsets of ±3 ppm with saturation powers ranging from 0-6 µT in 1 µT increments. The QUESP image acquisition parameters were: saturation pulse length = 3000 ms, TE/TR = 21/15000 ms, FA = 90°, matrix = 100x100, FOV = 30x30 mm, NA = 1.

T1 maps of the L-Arg phantoms were generated from variable repetition time (VTR) images acquired with repetition times TR = 7500, 5000, 3000, 1500, 800, 400, 200, and 50 ms. Phantom VTR image acquisition parameters were: TE = 6.5 ms, FA = 90°, matrix = 100x100, FOV = 30x30, NA = 1. For *in vivo* studies, T1 maps were generated from VTR images acquired with TR = 4000, 2000, 1500, 1000, 750, 400, and 100 ms. *In vivo* VTR image acquisition parameters were: TE = 7.5 ms, FA = 90°, matrix = 80x80, FOV = 40x40, NA = 1.

**MRF Dictionary Generation**

A large MRF dictionary of simulated signal intensity trajectories for a given acquisition schedule was generated using a custom-written MATLAB (Mathworks, Natick, MA) program. The dictionary simulations were performed using a vectorized, sparse matrix implementation of the Bloch equations modified to include chemical exchange between the water proton pool and both the solute (amide or amine) and semi-solid proton pools. Bloch equation simulations were performed for all possible combinations of a range of chemical exchange rates, exchangeable proton volume fractions, and water T1 and T2* relaxation times. The range of parameters for the phantom and *in vivo* studies are listed in Table 1. The *in vivo* water T1 and T2* relaxation times were fixed to the experimentally measured values. The value for the amine proton T2* of the L-Arg phantoms was selected from the best match between the experimental signal trajectory and a simulated MRF dictionary where the water and amine proton T1 were fixed and the amine T2* was varied from 20-60 ms in 10 ms increments. Generation of the ~670,000 dictionary entries (phantom experiments) required 61 minutes and 298 MB of storage on a 2.7 GHz Intel Core i7 MacBook Pro with 16 GB 1600 MHz DDR3 memory.

**Data Analysis**

For MRF experiments, the measured signal trajectories were normalized by their norms and matched voxelwise to the pre-computed CEST-MRF dictionary by selecting the entry with the largest vector dot product. For the QUESP experiments, the MTR$_{asym}$ at 3 ppm frequency offset was calculated and plotted as a function of saturation power. A custom-written MATLAB program was used to fit the saturation power curves for the exchange rate and water T1 and T2* using the lsqcurvefit function in MATLAB with the fit function defined by the Bloch equations. The L-Arg concentrations were kept fixed at the known concentrations. The 95% confidence intervals for the QUESP fit parameters were calculated from the residual and the jacobian matices using the nlparci MATLAB function. T1 maps were generated from an exponential fit of the variable TR signal intensity as a function of TR using a non-linear least squares fitting algorithm implemented in a custom-written MATLAB program. The Pearson correlation coefficients between the the MRF and QUESP determined chemical exchange rates, the MRF and known L-Arg concentrations, and the MRF and VTR determined T1 relaxation times were calculated in Prism 6 (GraphPad Software, Inc, La Jolla, CA). In addition, the pH dependence of the CEST-MRF determined exchange rate was fit to the acid/base catalyzed exchange model (38,39) as given by equation 3, where $k_0$ is the spontaneous exchange rate, $k_a$ is the acid catalyzed exchange rate, and $k_b$ is the base catalyzed rate.

$$k_{ex} = k_0 + \left(k_a \times 10^{-pH}\right) + \left(k_b \times 10^{pH-pKw}\right) \qquad \text{[Eq. 3]}$$

All statistical analyses were performed with Prism 6 (GraphPad Software, Inc, La Jolla, CA) with p<0.05 considered as significant. MRF matched parameter values for a given region-of-interest are reported as the mean ±standard deviation.

**Discrimination of CEST-MRF and CEST Z-spectrum acquisition schedules**

The capacities of two different MRF acquisition schedules – a variable saturation power schedule (used in this study) and a variable saturation frequency offset schedule (corresponding to a traditional CEST Z-spectrum) – to efficiently discriminate between different exchangeable proton concentrations and exchange rates were assessed by forming the dot-product correlation of the dictionaries with themselves, similar to previous work in the literature (40-42). Dictionary simulations for the variable saturation power acquisition schedule were performed using the L-Arg phantom acquisition schedule (Fig. 1A). Dictionary simulations for the CEST Z-spectrum schedule kept the saturation power fixed at 4 µT and varied the saturation frequency offset from +5 to -5 ppm in 0.25 ppm increments. Simulations were performed for five different exchangeable proton concentrations of 75, 150, 300, 600, and 900 mM with the exchange rate varied from 0-1000 Hz in 10 Hz increments. The low concentrations (75, 150, 300 mM) were chosen to match the range of L-Arg concentrations used experimentally (25, 50, 100 mM) as there are 3 equivalent $N_\alpha$-amine protons per L-Arg.

The discriminability of the water T1 relaxation times and exchangeable proton concentrations of the variable saturation power schedule was also calculated for the same five exchangeable proton concentrations with the water T1 varied from 2500-3300 ms in 20 ms increments.

**RESULTS**

**CEST-MRF of L-Arginine Phantoms**

CEST-MRF matched exchange rate, L-Arg concentration and water T1 maps are shown in Figure 2 for representative phantoms with either varying L-Arg concentration (Fig. 2, top row) or varying pH (Fig. 2, bottom row). For the L-Arg phantoms, there was a significant correlation (r=0.9964, p<0.0001) between the QUESP and CEST-MRF measured amine proton exchange rates (Fig. 3A). The pH dependence of the exchange rate (Fig. 3B) was well fit ($R^2$=0.9186) by the acid/base-catalyzed chemical exchange model given by equation 3 with $k_0$=252.2 Hz, $k_a$=1.42x10$^{-16}$ Hz, $k_b$=1.06x10$^{11}$ Hz, and p$K_w$=13.97, consistent with base catalyzed proton exchange.

A significant correlation (r=0.9526, p<0.0001) was also observed between the CEST-MRF and known L-Arg concentrations as shown in Figure 3C. The slight discrepancies in the matched L-Arg concentrations for the 100 mM L-Arg concentration phantoms are likely due to the relatively poor fingerprinting schedule efficiency as discussed below. While the CEST-MRF matched water T1 values were not significantly correlated (r=0.2207, p=0.4906) with the VTR measured T1 values (Fig. 3D), the CEST-MRF T1 values were all within ±20% of those measured by the VTR method. In general, the CEST-MRF T1 values tend to overestimate the VTR T1 values. A summary of the mean (±standard deviation) chemical exchange rates, L-Arg concentrations and water T1 relaxation times determined by the different measurement methods for regions-of-interest (ROIs) drawn for each phantom vial is given in Table 2.

**CEST-MRF of *In Vivo* Rat Brain**

CEST-MRF matched amide and semi-solid proton chemical exchange rates and exchangeable proton volume fraction maps for the *in vivo* rat brain along with the associated proton density image and Nissl stained rat brain section from the brainmap.org rat atlas (43) are shown in Figure 4. The average values for the matched parameters in white (corpus callosum and internal capsule) and gray (cerebral cortex) matter regions of interest are shown in Table 3. An average endogenous amide proton exchange rate of 36.6±12.8 Hz was measured in the rat brain cortex (Fig 4C) in good agreement with that measured previously (28.6±7.4 Hz) with the Water EXchange spectroscopy (WEX) method (44). The semi-solid proton pool volume fraction map demonstrated elevated volume fraction in white matter (11.2±1.7%) compared to gray matter (7.6±1.8%) brain regions (Fig 4E). The regions of elevated white matter semi-solid proton volume fraction are in good agreement with the Nissl stained histology tissue section (Fig. 4B), where neuronal cell bodies of gray matter, but not white matter fiber tracts, are preferentially stained.

**MRF Acquisition Schedule Efficiency**

The dictionary correlation matrices for the CEST-MRF and Z-spectrum acquisition methods are shown in Figure 5. For norm normalized signal trajectories, a vector dot product of unity represents a perfect match between two signal trajectories. For an ideal MRF acquisition schedule, a signal trajectory simulated for a particular exchange rate and proton volume fraction would be perfectly correlated only with itself (i.e., a vector dot product of one) and have a poor correlation (low vector dot product) with all other signal trajectories simulated for other combinations of proton exchange rate and volume fraction. The CEST-MRF correlation plots, however, demonstrate relatively poor discrimination of exchange rates, in particular at low exchangeable proton concentrations where trajectories with different exchange rates all have vector dot products very close to unity. Much better discrimination is, however, observed for the CEST-MRF acquisition schedule (Fig. 5A) than for the CEST Z-spectrum schedule (Fig. 5B). The discrimination of the water T1 relaxation times and exchangeable proton concentrations of the CEST-MRF acquisition schedule is shown in Figure 6. Very poor T1 discrimination is observed indicating that the CEST-MRF saturation power acquisition schedule is insensitive to T1 variations over the range of T1 values simulated (2500-3300 ms).

**DISCUSSION**

This study is the first application of MR Fingerprinting for the simultaneous quantification of proton chemical exchange rates and volume fractions. A previous study by Geades *et al* did use a pre-calculated look-up table of Bloch equation simulated Z-spectra to fit experimental Z-spectra, acquired with three different saturation powers, and extract proton volume fractions (45). However, due to the very large number of proton pools required to simulate the full Z-spectrum, a very coarse dictionary was used with only 8 different proton volume fractions for each of the NOE (aliphatic), CEST (amide) and MT (semi-solid) proton pools, 5 water T1 relaxation times and 3 B1 fields. All other parameters, including all chemical exchange rates, were fixed. Our CEST-MRF approach, in which the saturation frequency offset is fixed and the saturation power is varied provides several advantages over fitting or matching of traditional CEST Z-spectra. First, the CEST-MRF method is specific for the CEST (amide or amine) and MT (semi-solid) proton pools only. This greatly simplifies the analysis and allows finely sampled dictionaries to be used that can accurately quantify both the exchange rates and volume fractions of the CEST and MT exchangeable proton pools. Second, the experimental acquisition time is significantly reduced compared to a traditional CEST Z-spectrum. We have chosen an acquisition schedule with 30 iterations of the saturation power, which required less than 2 minutes to acquire. With further optimization of the acquisition schedule (41) even shorter acquisition times may be achievable. Third, the different exchangeable proton pools are sensitive to different RF saturation powers depending on their respective chemical exchange rates. Varying the RF saturation power provides simultaneous sensitivity to the various exchangeable proton pools. Fourth, as demonstrated by the dictionary correlation plots (Fig. 5), the CEST-MRF method provides greatly improved discrimination of proton exchange rates and volume fractions compared to a traditional CEST Z-spectrum. The very poor discrimination of the traditional CEST Z-spectrum may partly explain the very wide range of exchange rates and volume fractions that have been reported in the literature for the endogenous amide ($k_{ex}$=20-280 Hz, $f_s$=0.1-1.0%) and semi-solid ($k_{ex}$=1-160 Hz, $f_{ss}$=3-30.0%) proton pools of *in vivo* brain tissue.

While the CEST-MRF saturation power acquisition schedule displayed improved discrimination relative to the CEST Z-spectrum schedule, the schedule of saturation powers used in this work was selected at random and is likely far from optimal. This is reflected in the relatively poor discrimination of concentration and exchange rate observed in the correlation plots shown in Figure 5 implying strong similarity between signals arising from different tissue parameters. Nevertheless, excellent agreement was observed in L-Arg phantoms between the amine proton exchange rates and L-Arg concentrations calculated with MRF-CEST and those obtained with alternative techniques (Table 2). This can be attributed to the dictionary matching reconstruction since only the dictionary entry with the greatest dot product value is selected. Hence, acquisitions with poor discrimination can still yield accurate matches provided the signal-to-noise ratio (SNR) is sufficiently high to permit distinguishing similar signal evolutions. For a given SNR level, however, the discriminability can provide an *a priori* measure of the expected quality of the estimated tissue parameters.

The very poor CEST-MRF discrimination observed for the water T1 of the phantoms (Fig. 6) is consistent with the lack of a significant correlation between VTR and MRF determined T1. The lack of T1 sensitivity is not surprising for the CEST-MRF acquisition schedule used in this study, which only varied the saturation power and used a relatively long, constant repetition time (TR). This has the advantage that it simplifies the analysis, but at the cost of impaired T1 sensitivity. However, if accurate T1 maps are desired, sensitivity to T1 could be increased by varying additional acquisition parameters such as the TR and RF flip angle as originally demonstrated by Griswold and coworkers (35).

As discussed above, a wide range of endogenous amide proton exchange rates in brain tissue (20-280 Hz) have been reported in the literature (11,44-48). The water exchange spectroscopy (WEX) method should, however, provide the most accurate measurement of the amide proton exchange rate as the amide proton pool is specifically and directly probed (44) rather than being obtained from a large multiparametric fit of the CEST Z-spectrum. We allowed the amide proton exchange rate of the CEST-MRF dictionary to vary over a large range (5-150 Hz) encompassing most of the literature reported values and found good agreement between the *in vivo* amide proton exchange rate measured by CEST-MRF (36.6±12.8 Hz) and the previously reported water exchange (WEX) spectroscopy method (28.6±7.4 Hz) (44).

Similar good agreement was observed between our CEST-MRF matched semi-solid proton volume fractions ($f_{ss}$) in gray (7.6±1.8%) and white matter (11.2±1.7%) rat brain tissue and the values reported in MT studies (see Table 4) of rat (49-51), mouse (52,53), dog (54), bovine (55) and human brain tissue (56-59) with gray matter volume fractions ranging between 5-8% and white matter volume fractions between 10-15%. Variability in the semi-solid volume fractions was observed in human studies with some studies reporting significantly elevated gray (13%) and white (26%) matter semi-solid proton volume fractions (60,61), while others reported significantly lower gray (3-4%) and white matter (6-9%) volume fractions (45,48). However, in all cases an elevated semi-solid proton volume fraction was observed in white matter compared to gray matter brain tissue, consistent with the increased semi-solid lipid content of myelinated white matter fiber tracts. Previous MT studies have shown a strong correlation between $f_{ss}$ and histological measures of myelin fraction (49,52,62,63).

While the CEST-MRF saturation power acquisition schedule displayed relatively poor discrimination (Fig. 5), improved discrimination can be obtained by optimization of the acquisition schedule as was previously demonstrated for MRF (41) and multi-inversion EPI (40) sequences. The optimization consists of searching the hyperspace of acquisition parameters to find the set that maximizes the discrimination. Each of the acquisition parameters (saturation power, saturation frequency offset, saturation pulse length, TR, FA, etc.) represents an additional degree of freedom that can be used to improve the discrimination further. Importantly, improved discrimination can yield accurate estimation of the tissue parameters using only a small number of acquisitions, which translates directly into reduced acquisition times. This is an active area of research that will be explored in future studies. Although the $B_0$ field inhomogeneity was small for our study, with typical field shifts of only ±10 Hz, improved discrimination for studies with greater field inhomogeneity could be achieved by incorporating a range of $B_0$ field shifts into the simulated dictionary. This would allow for the simultaneous generation of exchange rate, proton volume fraction, and $B_0$ field maps, albeit at the cost of increased dictionary size that can, however, be mitigated using strategies described below.

An important challenge in the matching or optimization of CEST-MRF data is the large dictionaries that are required. This problem is particularly acute for *in vivo* acquisitions where the four-pool model contains up to fourteen nominally independent parameters hence theoretically requiring a fourteen dimensional dictionary. In this study a four dimensional dictionary with ~670,000 entries was used to limit the compute time needed. Nevertheless, larger dictionaries could provide improved accuracy and simultaneous estimation of additional parameters. Dictionary compression and fast matching methods (64,65) could be used to reduce the post-processing time. Unfortunately, those methods still require the full dictionary to first be generated prior to compression and may be ineffective on optimized acquisition schedules where similarities between the measured signals are minimized. Instead, recent work in Deep Learning based MRF reconstruction may be used to overcome this problem (66) through training of a neural network with sparse dictionaries that can be used to reconstruct the acquired CEST-MRF data.

**CONCLUSION**

CEST-MRF provides a method for fast, quantitative CEST imaging. Despite the relatively poor CEST-MRF discrimination, excellent agreement was observed between CEST-MRF and alternative methods for the proton exchange rates and volume fractions of both L-Arg phantoms and *in vivo* rat brain tissue. Optimization of the MRF acquisition schedule should lead to further improvements in the discrimination of chemical exchange rates and exchangeable proton volume fractions.

**TABLES**

**Table 1**

Range of tissue parameters used in each study denoted with the min:interval:max notation. The semi-solid tissue parameters were excluded from the phantom experiments since a two-pool model was used.

| Phantom | | *In Vivo* | |
|---|---|---|---|
| Amine exchange (Hz) | 100:10:1400 | Amide Exchange (Hz) | 5:5:150 |
| L-Arg concentration (mM) | 10:5:120 | Amide concentration (mM) | 100:50:1000 |
| Amine T1 (ms) | 2800 | Amide T1 (ms) | 1450 |
| Amine T2* (ms) | 40 | Amide T2* (ms) | 1 |
| Water T1 (ms) | 2500:50:3300 | Water T1 (ms) | 1450 |
| Water T2* (ms) | 600:50:1200 | Water T2* (ms) | 50 |
| | | Semi-solid exchange (Hz) | 5:5:100 |
| | | Semi-solid concentration (M) | 4:2:20 |
| | | Semi-solid T1 (ms) | 1450 |
| | | Semi-solid T2* (μs) | 40 |

**Table 2**

Comparison of the CEST-MRF determined L-Arg concentrations, amine proton chemical exchange rates and water T1 relaxation times with the known concentrations, QUESP measured exchange rates, and variable TR (VTR) measured T1 relaxation times, respectively, for the various L-Arg phantoms. The mean ±standard deviation was calculated for the ROI drawn around each vial.

| | [L-Arg] (mM) | | $k_{ex}$ (Hz) | | Water T1 (ms) | |
|---|---|---|---|---|---|---|
| pH | known | MRF | QUESP | MRF | VTR | MRF |
| 4.05 | 25 | 26.1±6.7 | 193.9±57.0 | 156.2±44.5 | 2492.8±51.8 | 2709.2±143.9 |
| 4.08 | 50 | 49.4±7.0 | 183.7±21.7 | 170.7±23.9 | 2802.9±60.0 | 2885.7±154.3 |
| 4.01 | 100 | 84.0±9.6 | 200.4±30.4 | 196.2±21.4 | 2628.3±52.6 | 2949±152.0 |
| 4.04 | 50 | 51.7±6.8 | 204.5±31.1 | 200.4±29.4 | 2669.7±55.7 | 2809.6±141.6 |
| 4.46 | 50 | 53.1±5.2 | 281.8±28.9 | 268.5±31.5 | 2608.3±50.2 | 2847.6±159.1 |
| 4.99 | 50 | 50.3±3.2 | 444.3±33.0 | 446.0±34.3 | 2854.1±58.7 | 2919.5176.2 |
| 5.05 | 25 | 26.9±2.7 | 384.2±61.1 | 375.5±53.6 | 2551.9±51.73 | 2500±0 |
| 5.02 | 50 | 55.1±3.4 | 420.9±45.3 | 392.4±39.5 | 2757.9±59.1 | 2507.7±34.6 |
| 5.02 | 100 | 110.0±6.4 | 437.8±82.7 | 424.5±30.7 | 2573.6±48.1 | 2721±197.3 |
| 4.99 | 50 | 50.4±2.9 | 444.1±86.7 | 491.5±40.9 | 2825.7±52.1 | 3073.2±244.6 |
| 5.43 | 50 | 62.7±3.6 | 670.0±120.8 | 692.6±50.6 | 2594.5±50.1 | 3117.4±229.3 |
| 5.96 | 50 | 57.9±3.4 | 1066.8±148.7 | 1219.4±65.0 | 2621.6±48.2 | 3164.4±222.3 |

**Table 3**

CEST-MRF determined amide and semi-solid proton chemical exchange rates ($k_{ex}$) and volume fractions ($f$) for gray (GM) and white matter (WM) regions of *in vivo* rat brain tissue.

|  | **Cortical GM** | **Subcortical WM**[a] |
|---|---|---|
| Amide $k_{ex}$ (Hz) | 36.6±12.8 | 49.1±12.9 |
| Amide $f_s$ (%) | 0.633±0.001 | 0.757±0.001 |
| Semi-solid $k_{ex}$ (Hz) | 46.0±6.5 | 47.7±2.9 |
| Semi-solid $f_{ss}$ (%) | 7.6±1.8 | 11.2±1.7 |

[a]average of corpus callosum and internal capsule

**Table 4**

Semi-solid, macromolecular proton volume fractions ($f_{ss}$) of gray (GM) and white matter (WM) brain tissue reported in the literature.

| Animal Species | GM $f_{ss}$ (%) | WM $f_{ss}$ (%) | Reference |
|---|---|---|---|
| rat | 7.6±1.8 | 11.2±1.7 | This study |
| rat | 6.6±0.2[a] | 12.9±0.4[b] | Underhill, *et al*, 2011 (49) |
| rat | 5.9±0.4 | 10.7±0.9 | Naumova, *et al*, 2017 (50) |
| rat | | 6.9±0.7[c] | Rausch, *et al*, 2009 (51) |
| mouse | 6.7±0.2 | 10.2±0.3 | Thiessen, *et al*, 2013 (52) |
| mouse (*ex vivo*) | 7.1±0.5 | 9.9±1.1 | Ou, *et al*, 2009 (53) |
| dog | 5.4±0.2 | 12.1±0.4 | Samsanov, *et al*, 2012 (54) |
| bovine (*ex vivo*) | 5.0±0.5 | 13.9±2.8 | Stanisz, *et al*, 2005 (55) |
| human | 6.4±1.2[d] | 15.7±2.4[e] | Sled, *et al*, 2001 (56) |
| human | 6.73±0.15 | 12.77±0.41 | Yarnykh, 2016 (57) |
| human | 5.77±0.34 | 13.48±0.37 | Yarnykh, *et al*, 2015 (58) |
| human | 7.60±1.17 | 14.29±1.08 | Yarnykh, 2012 (59) |
| human | 4.4±0.4 | 8.9±0.3 | Geades, *et al*, 2017 (45) |
| human | 3.4±0.4 | 6.2±0.4 | Liu, *et al*, 2013 (48) |
| human | | 25.0±1.2 | Jiang, *et al*, 2017 (61) |
| human | 13.6±0.8[f] | 26.2±1.2 | van Gelderen, *et al*, 2017 (60) |

[a]average of caudate, cerebellar GM, cortical GM, superior colliculus and thalamus. [b]average of anterior limb/anterior commissure, corpus callosum, internal capsule and ventral hippocampal WM. [c]average of corpus callosum and medulla. [d]average of cortex and caudate nucleus. [e]average of frontal WM from 2 subjects. [f]average of globus pallidus, putamen and caudate nucleus.

**FIGURES**

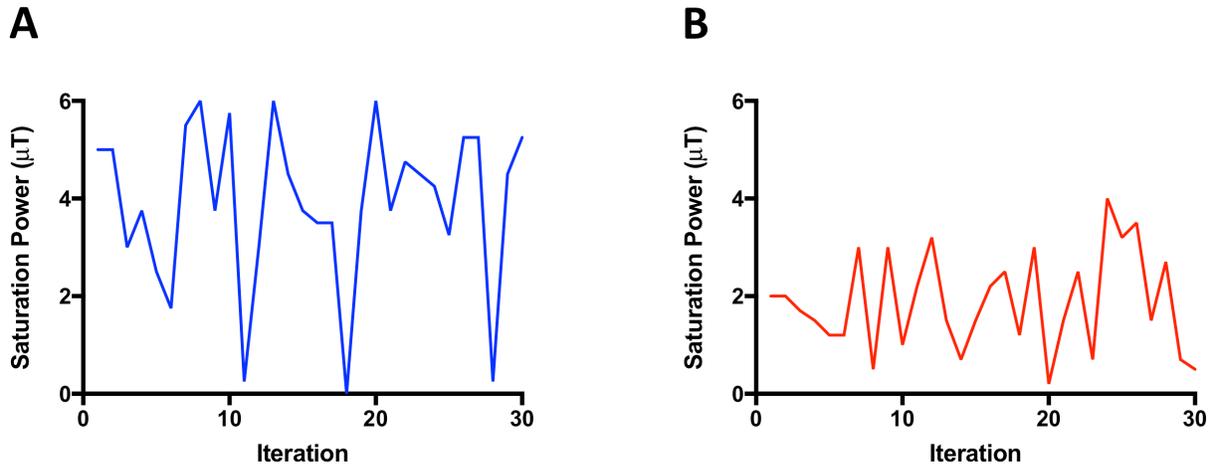

**FIG. 1.** Schematic of the CEST-MRF acquisition schedule. The saturation pulse power was varied for 30 iterations ranging from (A) 0-6 µT for the L-Arginine phantoms and (B) 0-4 µT for the *in vivo* rat brain. For the phantom study a 3 sec saturation pulse was used with the saturation frequency offset fixed at 3 ppm, corresponding to the frequency of the $N_\alpha$-amine exchangeable protons. For the *in vivo* study, a 2.5 sec saturation pulse was used with the saturation frequency offset fixed at 3.5 ppm, corresponding to the frequency of the amide exchangeable protons.

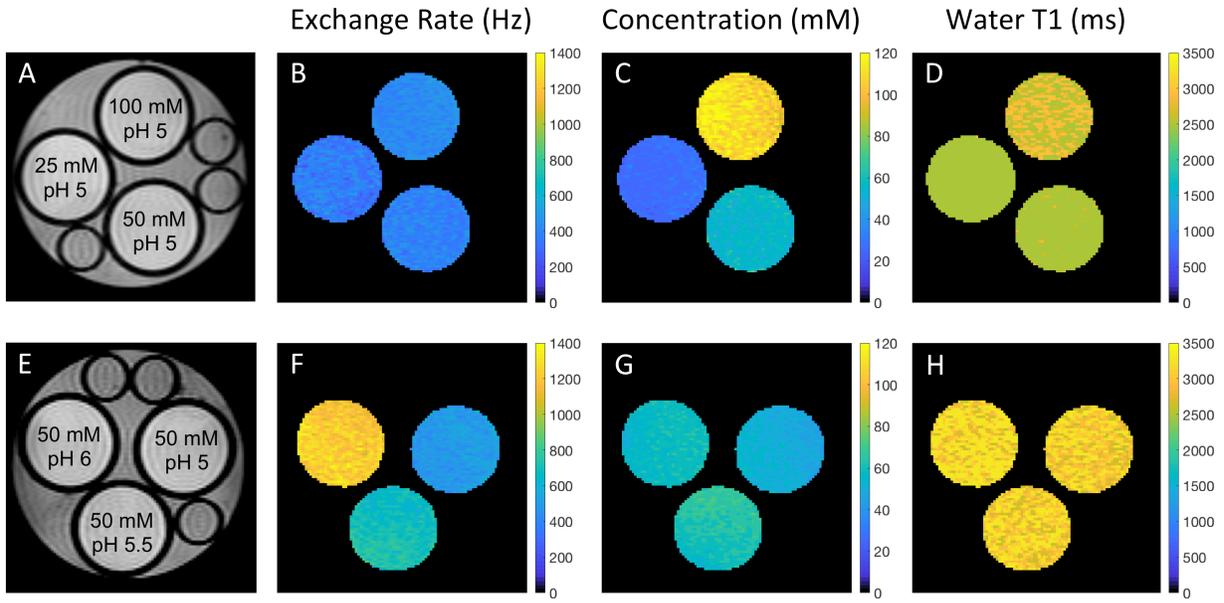

**FIG. 2.** (A,E) Proton density images of representative L-arginine phantoms with varying concentrations (top row) and pH (bottom row) along with the associated quantitative chemical exchange rate (B,F), L-arginine concentration (C,G), and water T1 (D,H) maps generated from the MRF matching.

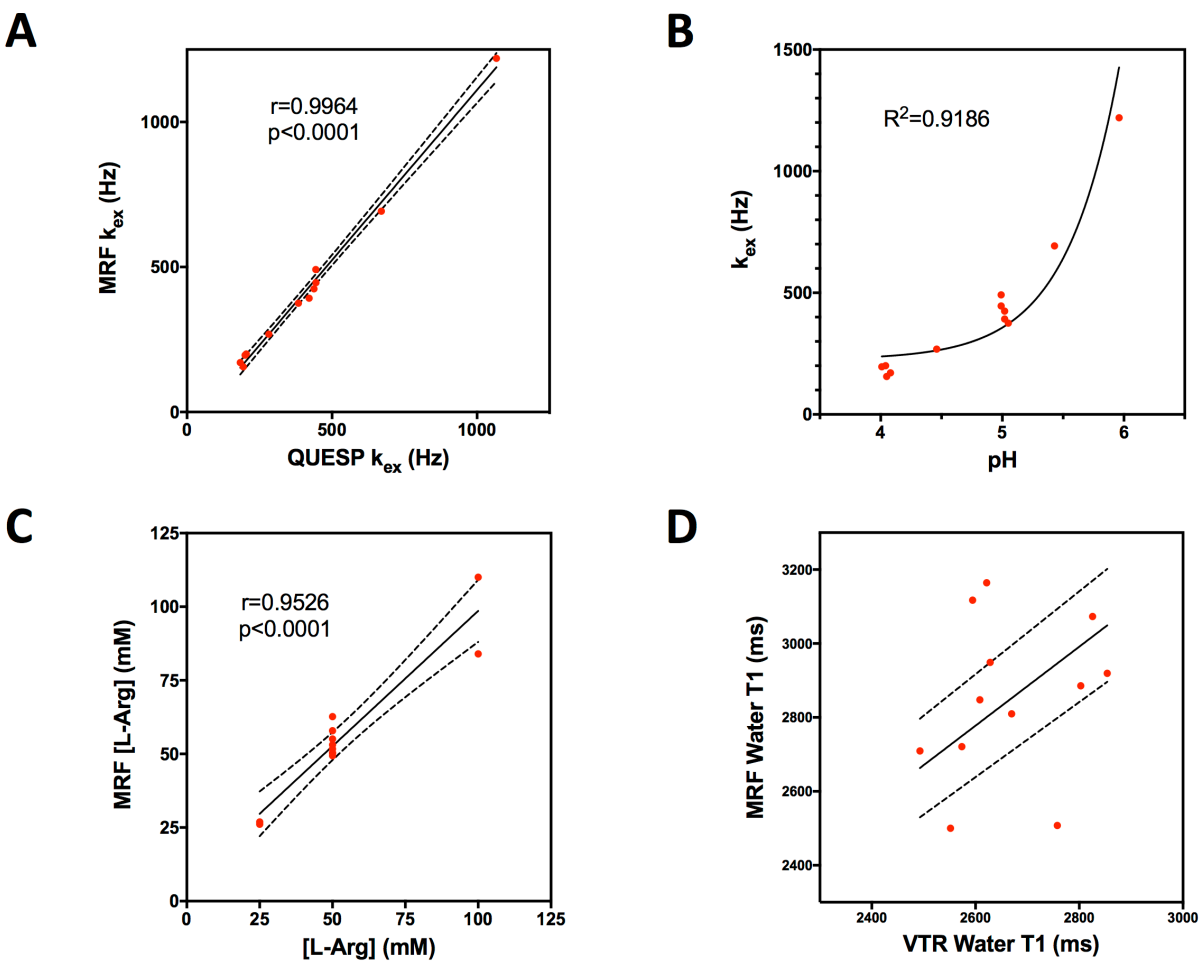

**FIG. 3.** (A) The MRF determined exchange rates for the $N_\alpha$-amine protons of L-Arg were significantly correlated (r=0.9964, p<0.0001) with the exchange rates determined with the QUESP method. (B) The pH dependence of the chemical exchange rate was well fit ($R^2$=0.9186) by the base catalyzed proton exchange model (Eq. 3). (C) The MRF determined L-Arg concentrations were significantly correlated (r=0.9526, p<0.0001) with the known concentrations. (D) The water T1 relaxation times measured by MRF and a variable repetition time (VTR) method were not significantly correlated, however, the MRF determined relaxation times were all within ±20% of the VTR values. The dashed lines represent the 95% confidence interval.

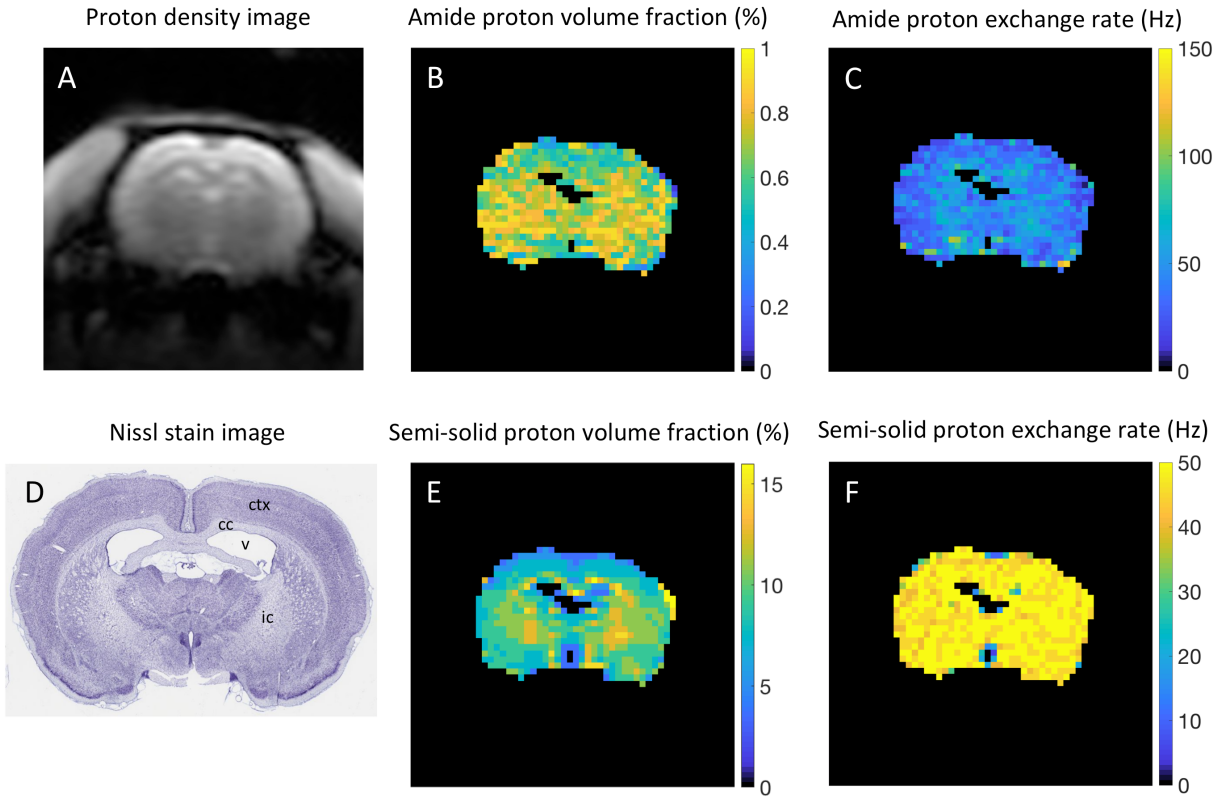

**FIG. 4.** (A) Proton density image and (D) corresponding Nissl stained rat brain section with the ventricle (v), cortex (ctx), corpus callosum (cc), and internal capsule (ic) identified. MRF matched (B) amide proton volume fraction and (C) chemical exchange rate maps and semi-solid (E) proton volume fraction and (F) chemical exchange rate maps of *in vivo* rat brain tissue. The average cortex amide proton exchange rate determined from MRF (36.6±12.8 Hz) was in good agreement with that measured previously using the water exchange spectroscopy (WEX) method (28.6±7.4 Hz). The semi-solid proton volume fraction was elevated in white (11.2±1.7%) compared to gray (7.6±1.8%) matter brain regions in agreement with literature values.

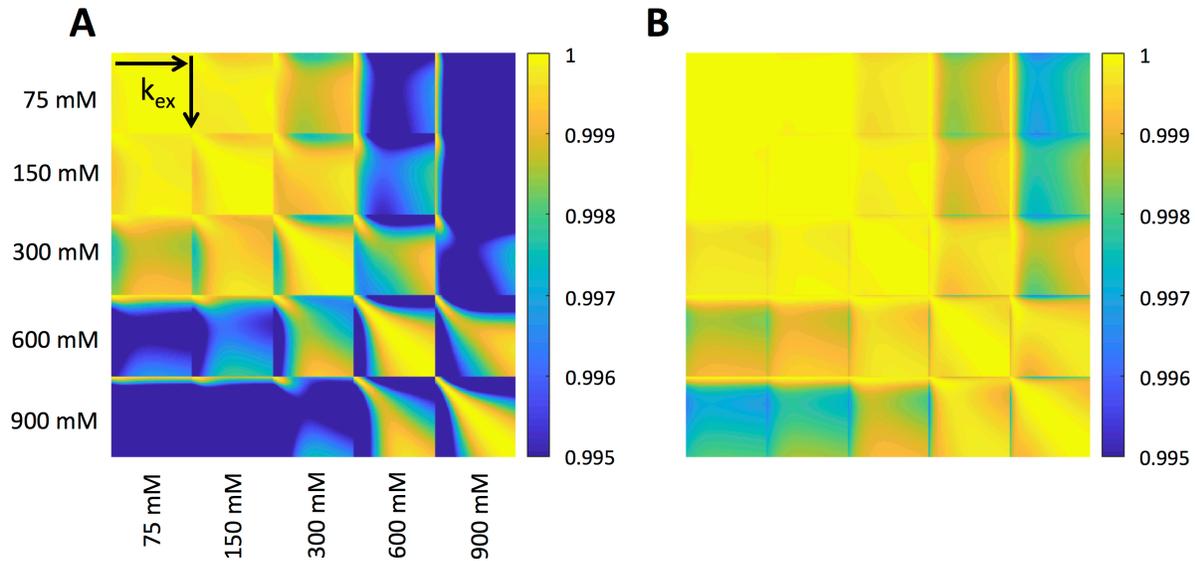

**FIG. 5.** Correlation of the MRF simulated dictionary with itself for MRF acquisition schedules that varied either (A) the saturation pulse power or (B) the saturation pulse frequency offset, corresponding to a traditional CEST Z-spectrum. The correlation was performed for 5 different exchangeable proton concentrations (75, 150, 300, 600, and 900 mM) with the chemical exchange rate varied for each concentration from 0-1000 Hz in 10 Hz increments. Better chemical exchange rate and concentration discrimination was observed with the variable saturation power acquisition schedule than with the CEST Z-spectrum acquisition schedule.

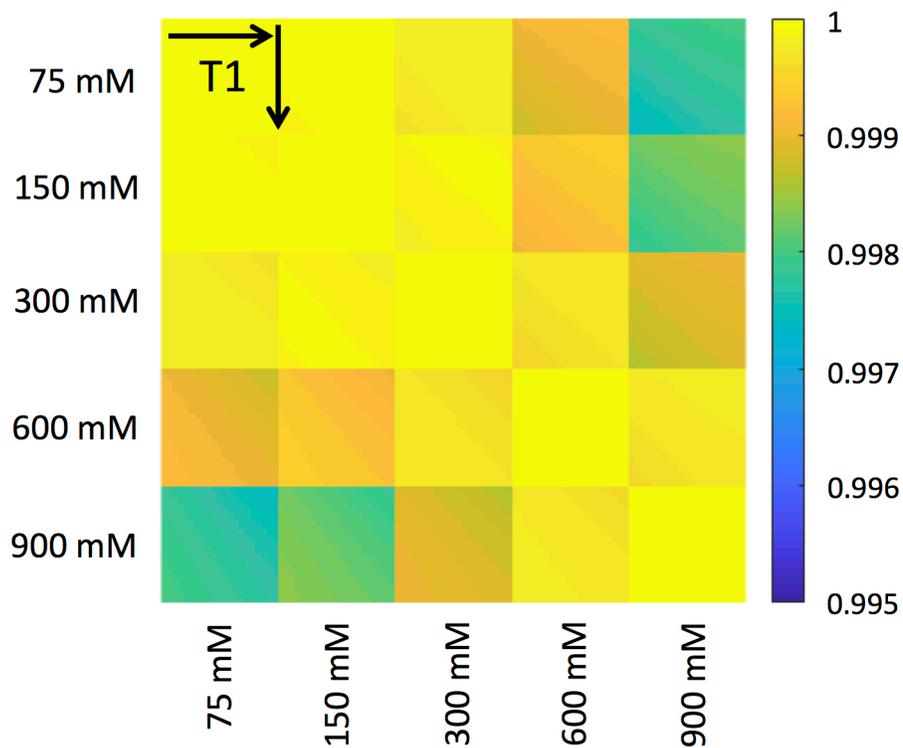

**FIG. 6.** Correlation of the MRF simulated dictionary with itself for the variable saturation power acquisition schedule. The correlation was performed for 5 different exchangeable proton concentrations (75, 150, 300, 600, and 900 mM) with the water T1 relaxation time varied for each concentration from 2500-3200 ms in 20 ms increments. The variable saturation power acquisition schedule demonstrated little sensitivity to the water T1.